\documentclass[a4paper,11pt]{article}
\pdfoutput=1 

\usepackage{jcappub} 

\usepackage[T1]{fontenc}
\usepackage{mathrsfs}
\usepackage{amsmath}
\usepackage{upgreek}

\title{\boldmath Reshaping the inner shadow of a Kerr black hole by a torn accretion disk}

\author[a,1]{Shiyang Hu,\note{Corresponding author.}}
\author[a]{Dan Li,}
\author[b]{Chen Deng,}
\author[c]{and Kejian He}

\affiliation[a]{School of Mathematics and Physics, University of South China, \\ Hengyang, 421001 People's Republic of China}
\affiliation[b]{School of Astronomy and Space Science, Nanjing University, \\ Nanjing, 210023 People's Republic of China}
\affiliation[c]{Department of Mechanics, Chongqing Jiaotong University, \\ Chongqing, 400000 People's Republic of China}

\emailAdd{husy\_arcturus@163.com}
\emailAdd{danli@usc.edu.cn}
\emailAdd{dengchen@smail.nju.edu.cn}
\emailAdd{kjhe94@163.com}

\abstract{When an accretion flow extends to the event horizon, their intersection defines the contour of the inner shadow. However, the morphological evolution of this critical feature remains largely unexplored within a torn accretion disk system, a configuration comprising distinct sub-disks formed when a tilted disk is disrupted by frame-dragging. To address this, we phenomenologically construct a torn accretion disk model and numerically simulate the inner shadow of a Kerr black hole using relativistic backward ray-tracing. We discover that the torn disk geometry profoundly alters the black hole's observational signatures, inducing severe erosion of the inner shadow and generating novel features such as bifurcated shadows, crescent-like structures, and multiple orders of shadow rings. These exotic morphologies, which are predominantly governed by the spatial discontinuity between the sub-disks and the tilt angle of the outer sub-disk, are exceedingly difficult to replicate within standard equatorial accretion paradigms. Our findings demonstrate that these distinctive shadow structures hold significant potential to serve as robust diagnostic probes for torn accretion environments, simultaneously implying that relying solely on the inner shadow to test gravity theories is fundamentally insufficient.}

\begin{document}
\maketitle
\flushbottom

\section{Introduction}
In recent years, the Event Horizon Telescope (EHT) collaboration has achieved an unprecedented angular resolution of $20-25$ $\mu$as using very long baseline interferometry (VLBI), transforming the theoretical prospects of observing black hole shadows proposed in the early 2000s \cite{2000ApJ...528L..13F} into observational reality \cite{2019ApJ...875L...1E,2022ApJ...930L..12E}. The images of the supermassive black holes M87$^{*}$ and Sagittarius A$^{*}$ reveal a central dark depression surrounded by an asymmetric bright emission ring. While the macroscopic morphology of this bright ring is primarily governed by the complex radiation of the surrounding plasma, its inner edge is closely associated with the boundary of the black hole shadow, which is theoretically defined by photons on critical unstable bound orbits. Because the geometric size and shape of the shadow are predominantly determined by the background spacetime metric and remain largely independent of the specific astrophysical details of the accretion flow \cite{2019ApJ...885L..33N,2021ApJ...920..155B}, this structure has emerged as a powerful tool for testing general relativity and alternative gravity theories in the strong-field regime \cite{2015ApJ...814..115P,2016PhRvL.116c1101J,2018CQGra..35w5002A,2018JCAP...07..015H,Guo:2018kis,2020EPJC...80..872Z,2021EPJP..136..436W,2022PhR...947....1P,2022ApJ...932...51A,2023IJMPD..3250088H,2025arXiv251106017L}.

Notwithstanding the landmark advancements made by the EHT, current angular resolution remains insufficient to resolve the fine structures of black hole images, such as the sharp boundary of the shadow or the precise location of the event horizon \cite{2019PhRvD.100b4018G}. This limitation leaves substantial room for alternative interpretations within modified gravity theories \cite{2015PhRvL.115u1102C,2018NatAs...2..585M,Peng:2020wun,2021PhRvD.104h1503G,Hou:2021okc,2023PhRvD.107f4001G,Wang:2023nwd,Zhang:2023bzv,Wang:2025vsx,Wan:2025gbm,Aslam:2025hgl,Zeng:2025kyv,Zeng:2025tji,Zeng:2025kqw,He:2025hbu,He:2024amh,Zhao:2025nwi,Zheng:2024ftk,Huang:2025xqd,Yang:2026bgi,He:2026odk,Zeng:2026pch}. To bridge this gap, the proposed next-generation EHT (ngEHT) project plans to deploy space-based interferometers and operate them in conjunction with the ground-based EHT. By significantly extending the interferometric baselines, this initiative aims to achieve multi-frequency, high-resolution black hole imaging \cite{2021MNRAS.500.4866A,2022MNRAS.511..668L,2022Galax..10..111T,2023Galax..11...61J,2023Galax..11...15R,2025LRR....28....4A}. Consequently, the ngEHT aims to reveal more intricate features, specifically the signature of the event horizon itself. Notably, when the inner edge of an accretion disk extends toward the event horizon, their intersection creates a closed dark region on the observational screen that is smaller than the shadow cast by the critical photon orbit. This feature, originally termed the ``central shadow'' \cite{2003PASJ...55..155F}, and whose morphology was subsequently further investigated by \cite{2004ApJ...611..996T,2019JETP..128..578D}, has now become widely recognized as the ``inner shadow'' \cite{2021ApJ...918....6C,Hou:2022eev,Wang:2023fge,2024JCAP...04..089H,Hou:2024qqo,2025JCAP...06..036H,2025arXiv251106219W,2026ForPh..7470062L,2026arXiv260407726C}. As a definitive signature of the event horizon and a powerful probe for the intrinsic black hole parameters, investigating the inner shadow has become a critical scientific objective for the ngEHT.

Based on time-averaged images derived from 3D general relativistic magnetohydrodynamic (GRMHD) simulations, Chael et al. constructed an analytic equatorial accretion disk model for Kerr black holes, numerically simulating the corresponding inner shadows at $230$ GHz and $86$ GHz \cite{2021ApJ...918....6C}. They demonstrated that the inner shadow contour represents the direct projection of the intersection between the event horizon and the disk's inner edge onto the observer's screen, essentially acting as a lensed image of the event horizon. Furthermore, they explored the influence of the inclination angle and spin parameter on the shadow's geometry, highlighting the feasibility of jointly constraining the black hole spin and mass-to-distance ratio by combining the average radii of the inner shadow and the critical curve \cite{2019PhRvD.100b4018G}. Expanding this paradigm, Hou et al. \cite{Hou:2022eev} employed ray-tracing algorithms \cite{Hu:2020usx,Zhong:2021mty} to simulate a rotating black hole illuminated by an optically and geometrically thin equatorial disk within the Melvin universe. They found that the ambient magnetic field modifies the inner shadow's shape; specifically, an enhanced magnetic field elongates the shadow horizontally and compresses it vertically, revealing its potential as a robust probe for external magnetic environments. Additionally, they noted that the geometric properties of the inner shadow respond to black hole parameters in a manner consistent with the traditional shadow boundary (the critical curve), implying that the inner shadow can serve as a spacetime fingerprint under equatorial disk assumptions. Interestingly, Wang et al. discovered that within specific parameter spaces, the inner shadow of a Kerr-Taub-NUT black hole exhibits a distinctive ``duck-cap-like'' morphology, which can act as a unique identifier for this spacetime \cite{2025arXiv251106219W}. Collectively, these findings point to a clear conclusion: in the presence of a geometrically thin equatorial accretion disk, a robust mapping exists between the inner shadow and the background spacetime.

However, recent investigations have introduced a pivotal shift in this paradigm. Within the framework of Horndeski gravity, some authors of the present work numerically simulated the images of spherically symmetric, scalar-hairy black holes surrounded by tilted accretion disks \cite{2024JCAP...04..089H}. This study revealed that the morphology of the inner shadow is highly sensitive to the disk's inclination: specifically, a tilted disk possesses the capacity to erode the inner shadow. Furthermore, they demonstrated that the precession of a tilted disk induces a corresponding rotation of the inner shadow. These findings imply that the inner shadow is intrinsically dependent on the accretion environment. Admittedly, the analysis in \cite{2024JCAP...04..089H} was restricted to spherically symmetric black holes. This limitation renders the model somewhat inadequate, given the astrophysical consensus that realistic black holes possess non-zero spin. To address this gap, Hu et al. further investigated the inner shadows of Kerr black holes illuminated by tilted, geometrically thin accretion disks \cite{2025JCAP...06..036H}. They noted that under certain conditions, such as a substantial disk inclination, a Kerr black hole can exhibit a vanishingly small inner shadow. The area of this shadow can be significantly smaller than the theoretical minimum allowable in equatorial disk models\footnote{For an equatorial accretion disk, the inner shadow of a Kerr black hole decreases in size with an increasing spin parameter and enlarges with a higher viewing angle. Consequently, the Kerr black hole exhibits its minimum inner shadow when the viewing angle is $0^{\circ}$ and the dimensionless spin parameter approaches $1$ \cite{2025JCAP...06..036H}.}. Moreover, they found that for specific combinations of the spin parameter and disk tilt angle, the inner shadow of a Kerr black hole splits into a ``primary shadow'' and a ``secondary shadow''. The former generally maintains an elliptical shape, while the latter frequently appears eyebrow-like. These discoveries reaffirm that the inner shadow is inextricably linked to the accretion environment, cautioning that utilizing it as a direct test for black hole theories requires meticulous care.

Given that the geometric properties of the inner shadow are highly sensitive to the accretion environment, it is imperative to expand the theoretical template library of inner shadows across a broader spectrum of accretion conditions. Intriguingly, through numerical simulations, several authors demonstrated that for rapidly spinning Kerr black holes, a tilted, relatively thin accretion disk can be torn into multiple distinct sub-disks \cite{2012ApJ...757L..24N,2021MNRAS.507..983L,2023ApJ...944L..48L,2023MNRAS.518.1656M,2023ApJ...955...72K,2024JCAP...07..063S,2025arXiv251109626K}. This tearing phenomenon arises because the differential torques induced by frame-dragging exceed the viscous torques that hold the disk together \cite{2021MNRAS.507..983L}. Subsequently, driven by the Bardeen-Petterson effect \cite{2021MNRAS.507..983L,1975ApJ...195L..65B,2013Sci...339...49M,2015MNRAS.448.1526N}, the angular momentum of the inner sub-disk aligns with the black hole spin on a short timescale, whereas the outer sub-disk essentially retains its initial inclination over a substantially longer period. Consequently, over astrophysically relevant timescales, the Kerr black hole hosts a torn accretion disk system composed of an aligned inner equatorial disk and a misaligned outer tilted disk. The specific impact of this complex accretion architecture on the inner shadow warrants rigorous investigation, which directly forms the primary motivation of this work.

In this paper, adopting a purely geometric perspective, we construct a torn accretion disk model and numerically simulate the inner shadow of a Kerr black hole across an extensive parameter space to unveil novel observational signatures. The remainder of this paper is organized as follows. In section 2, we briefly review the photon geodesic equations in Kerr spacetime, laying the groundwork for the accretion model and ray-tracing methodology introduced in section 3. In section 4, we present the morphological evolution of the Kerr black hole inner shadow under the combined influence of the viewing inclination, disk tilt angle, and the disk tearing radius, thereby revealing new geometric structures of the inner shadow. Finally, in section 5, we summarize our results and provide concluding remarks. Throughout this paper, we adopt geometric units ($G=c=M=1$).
\section{Kerr metric and null-like geodesics}
In the Boyer-Lindquist coordinates $x^{\mu} = (t,r,\theta,\varphi)$ with the metric signature $(-,+,+,+)$, the covariant metric components of the Kerr black hole are \cite{Cunha:2016bpi,2021ApJ...914...63W,2021EPJC...81..785S}
\begin{eqnarray}
g_{tt} &=& -\left(1-\frac{2r}{\Sigma}\right), \label{1} \\
g_{rr} &=& \frac{\Sigma}{\Delta}, \label{2} \\
g_{\theta\theta} &=& \Sigma, \label{3} \\
g_{\varphi\varphi} &=& \left(r^{2}+a^{2}+\frac{2ra^{2}\sin^{2}\theta}{\Sigma}\right)\sin^{2}\theta, \nonumber \\
g_{t\varphi} &=& g_{\varphi t} = -\frac{2ar\sin^{2}\theta}{\Sigma}, \label{4}
\end{eqnarray}
where $a$ denotes the dimensionless spin parameter, which is fixed at $a=0.94$ in our subsequent calculations. The functions $\Sigma$ and $\Delta$ are defined as $\Sigma = r^{2}+a^{2}\cos^{2}\theta$ and $\Delta = r^{2}+a^{2}-2r$, respectively. By solving the equation $\Delta = 0$, the event horizon radius is determined to be $r_{\textrm{e}} \approx 1.34$. This radius not only marks the location of the one-way membrane of the black hole but also serves as the inner boundary of the accretion disk in our model.

According to the covariant metric, the Lagrangian $\mathscr{L}$ governing particle motion in Kerr spacetime is formulated as
\begin{eqnarray}\label{5}
\mathscr{L} = \frac{1}{2}g_{\mu\nu}\dot{x}^{\mu}\dot{x}^{\nu},
\end{eqnarray}
where $\dot{x}^{\mu}=(\dot{t},\dot{r},\dot{\theta},\dot{\varphi})$ represents the four-velocity of the particle. We further introduce the conjugate momentum $p_{\mu}$, which is related to the four-velocity via $p_{\mu}=\partial \mathscr{L} / \partial \dot{x}^{\mu}$:
\begin{eqnarray}
p_{t} &=& g_{tt}\dot{t} + g_{t\varphi}\dot{\varphi} = -\left(1-\frac{2r}{\Sigma}\right)\dot{t}-\frac{2ar\sin^{2}\theta}{\Sigma}\dot{\varphi}, \label{6} \\
p_{r} &=& g_{rr}\dot{r} = \frac{\Sigma}{\Delta}\dot{r}, \label{7} \\
p_{\theta} &=& g_{\theta\theta}\dot{\theta} = \Sigma \dot{\theta}, \label{8} \\
p_{\varphi} &=& g_{\varphi\varphi}\dot{\varphi} + g_{t\varphi}\dot{t} = \sin^{2}\theta\left(r^{2}+a^{2}+\frac{2a^{2}r\sin^{2}\theta}{\Sigma}\right)\dot{\varphi}-\frac{2ar\sin^{2}\theta}{\Sigma}\dot{t}. \label{9}
\end{eqnarray}
Given the stationary and axisymmetric properties of the spacetime, the Lagrangian $\mathscr{L}$ is independent of the coordinates $t$ and $\varphi$. Consequently, the conjugate momenta $p_{t}$ and $p_{\varphi}$ are conserved quantities, corresponding to the negative specific energy and the specific angular momentum of the particle under the adopted metric signature, denoted as $p_{t}=-E$ and $p_{\varphi}=L$, respectively.

Subsequently, by applying the Legendre transformation, the Hamiltonian $\mathscr{H}$ governing particle motion in the Kerr spacetime is expressed as
\begin{equation}\label{10}
\mathscr{H}= p_{\mu}\dot{x}^{\mu}-\mathscr{L} = \frac{1}{2}g^{\mu\nu}p_{\mu}p_{\nu}.
\end{equation}
Here, $g^{\mu\nu}$ denotes the contravariant metric components, which take the form \cite{2021ApJ...914...63W,2021EPJC...81..785S}
\begin{eqnarray}
g^{tt} &=& -\frac{\left(r^{2}+a^{2}\right)^{2}-\Delta a^{2}\sin^{2}\theta}{\Delta\Sigma}, \label{11} \\
g^{rr} &=& \frac{\Delta}{\Sigma}, \label{12} \\
g^{\theta\theta} &=& \frac{1}{\Sigma}, \label{13} \\
g^{\varphi\varphi} &=& \frac{\Sigma-2r}{\Delta\Sigma\sin^{2}\theta}, \label{14} \\
g^{t\varphi} &=& g^{\varphi t} = -\frac{2ar}{\Delta\Sigma}. \label{15}
\end{eqnarray}
By substituting these components along with the conserved quantities, the explicit expression for the Hamiltonian is obtained as
\begin{eqnarray}\label{16}
\mathscr{H} =\frac{1}{2}\left[-\frac{\left(\Delta + 2r\right)^{2}-\Delta a^{2}\sin^{2}\theta}{\Delta\Sigma}(-E)^{2}+\frac{\Delta}{\Sigma}p_{r}^{2}+\frac{p_{\theta}^{2}}{\Sigma}+\frac{\Sigma-2r}{\Delta \Sigma \sin^{2}\theta}L^{2}-\frac{4ar}{\Delta \Sigma}(-E)L\right].
\end{eqnarray}

For photons propagating in the Kerr spacetime, the null geodesic equations are fundamentally governed by Hamilton's canonical equations, given by
\begin{equation}\label{17}
\dot{x^{\mu}}=\frac{\partial \mathscr{H}}{\partial p_{\mu}}, \quad \dot{p_{\mu}} = -\frac{\partial \mathscr{H}}{\partial x^{\mu}}.
\end{equation}
Once the initial coordinates $x^{\mu} = (t,r,\theta,\varphi)$ and the corresponding conjugate momenta $p_{\mu}=(p_{t},p_{r},p_{\theta},p_{\varphi})$ are specified, the propagation trajectory of the photon is uniquely determined. Furthermore, throughout the entire motion, the photon four-momentum strictly satisfies the Hamiltonian constraint for null geodesics, namely $\mathscr{H}=0$.
\section{Torn disk and imaging process}
A tilted accretion disk surrounding a rotating black hole can be torn into distinct inner and outer sub-disks driven by frame-dragging effects. Governed by the Bardeen-Petterson effect, the angular momentum of the inner disk rapidly aligns with the black hole's spin axis, whereas the outer disk stably maintains its original inclination over a relatively long period. From a purely geometric perspective, this astrophysical scenario is phenomenologically illustrated in figure 1. The local coordinate system of the black hole is denoted by $(x^{\prime},y^{\prime},z^{\prime})$, with the black hole situated at the origin. The torn accretion disk system consists of two fundamental components: an outer tilted sub-disk, which subtends an angle $\sigma$ with respect to the equatorial plane, and an inner sub-disk located entirely within the equatorial plane. The radial coordinate where the two sub-disks connect is denoted by $r_{\textrm{cut}}$, defining the tearing radius of the accretion disk. Furthermore, it is crucial to emphasize that we can also assume a radial discontinuity after the disk tears. Specifically, the inner boundary of the outer tilted disk, $r^{\textrm{outer}}_{\textrm{in}}$, may not equal the outer boundary of the inner equatorial disk, denoted by $r^{\textrm{inner}}_{\textrm{out}}$ ($r^{\textrm{outer}}_{\textrm{in}} \neq r^{\textrm{inner}}_{\textrm{out}}$). Consequently, the free parameters characterizing the phenomenological torn disk model include $\sigma$, $r_{\textrm{cut}}$, or alternatively $r^{\textrm{outer}}_{\textrm{in}}$ and $r^{\textrm{inner}}_{\textrm{out}}$. We artificially color-code the two disk components to intuitively visualize the final image and explicitly trace the origin of the observed emission.
\begin{figure*}
\center{
\includegraphics[scale=0.7]{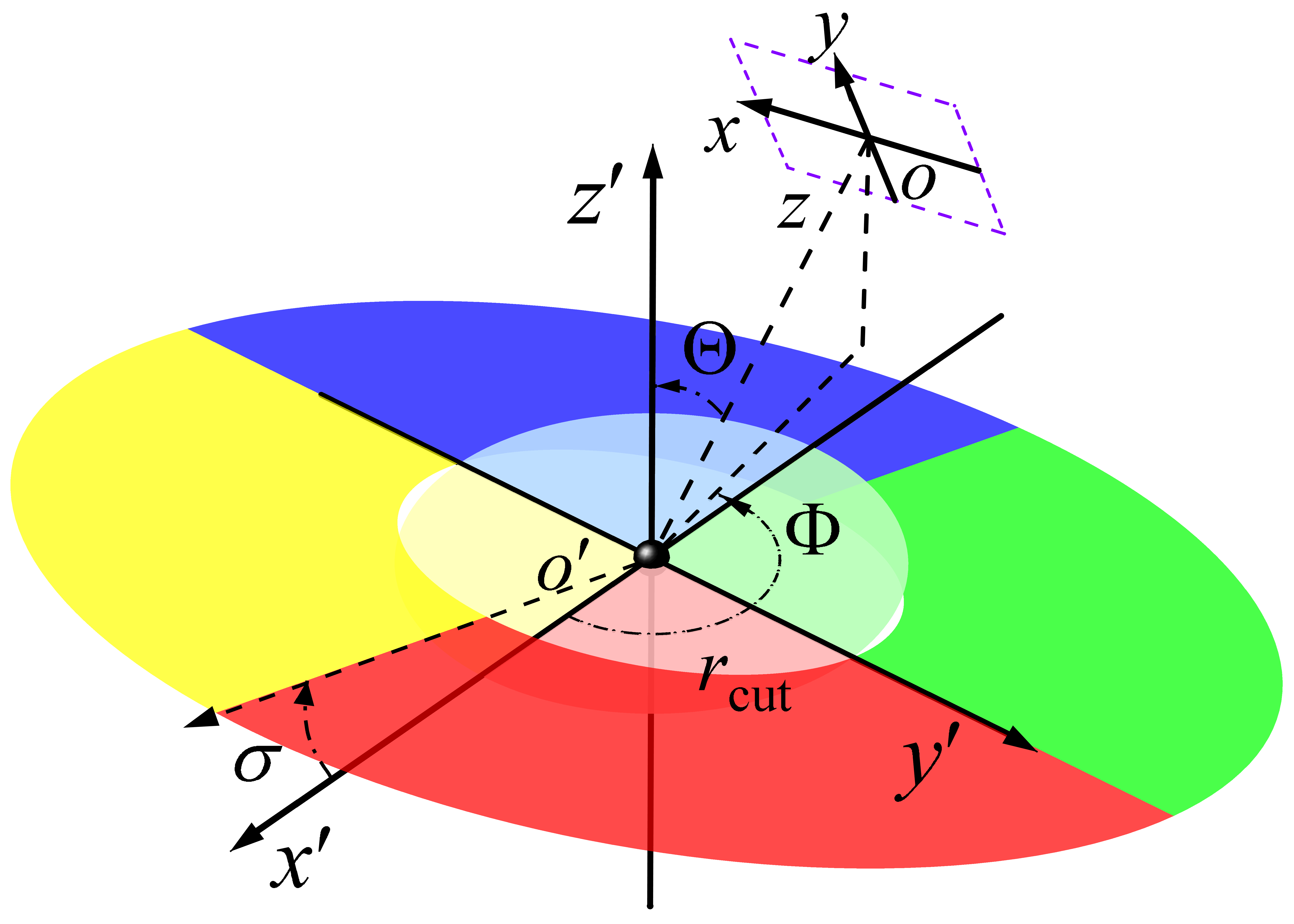}
\caption{Schematic illustration of the torn accretion disk and the associated coordinate systems. In the local frame of the black hole $(x^{\prime},y^{\prime},z^{\prime})$, the black hole is located at the origin, with the $o^{\prime}z^{\prime}$ axis aligned with its spin axis. The torn accretion disk comprises an inner and an outer region. The outer tilted sub-disk maintains an annular configuration with an inclination angle $\sigma$. The inner boundary of the inner equatorial sub-disk intersects the black hole event horizon. When the two sub-disks are radially continuous, the transition is characterized by the tearing radius $r_{\textrm{cut}}$. When they are discontinuous, the outer boundary of the inner disk is denoted by $r^{\textrm{inner}}_{\textrm{out}}$, and the inner boundary of the outer tilted disk is denoted by $r^{\textrm{outer}}_{\textrm{in}}$. Distinct colors are assigned to the two sub-disk components to facilitate the visualization of the black hole image. The observer's local coordinate frame is defined as $(x,y,z)$. The viewing inclination, denoted by $\Theta$, is defined as the angle between the line of sight $o^{\prime}o$ and the black hole spin axis. The observer's azimuthal angle, $\Phi$, is defined as the angle between the projection of $o^{\prime}o$ onto the black hole equatorial plane and the $x^{\prime}$ axis.}} \label{fig1}
\end{figure*}

The torn accretion disk serves as the primary light source illuminating the black hole. From a realistic astrophysical perspective, a portion of the emitted electromagnetic radiation is captured by a distant observer, while the remainder plunges into the black hole, contributing to the shadow on the observational screen. This imaging logic can be effectively implemented utilizing the relativistic backward ray-tracing technique \cite{Hu:2020usx,Zhong:2021mty,Cunha:2016bpi,2010ApJ...718..446J,2011CQGra..28v5011V,2012ApJ...745....1P,2012ApJ...761..174B,2012A&A...545A..13Y,2013ApJ...762..122C,2013ApJS..207....6Y,2015CQGra..32f5001J,2016ApJ...820..105P,2022EPJC...82..103V,2023CQGra..40m5011G,2024JCAP...11..054H,2022MNRAS.517.3711T,2024CQGra..41i5010A,2025arXiv251012585H,Boos:2025nzc}. As illustrated in figure 1, the spatial region in the vicinity of an observer with a viewing inclination $\Theta$ and an azimuthal angle $\Phi$ is described by a local coordinate frame $(x,y,z)$. Here, the $\overline{xoy}$ plane constitutes the observational screen, with $x$ and $y$ acting as the screen coordinates. Given that the observation distance $r_{\textrm{obs}}$ (the distance $oo^{\prime}$) is assumed to be sufficiently large, the light rays emitted from the accretion disk and received by the observer can be regarded as striking the screen perpendicularly along the negative $z$-direction. 

By treating each pixel $(x,y)$ on the screen as the initial position, we trace the photon trajectories backward in time to determine whether they intersect the black hole event horizon or the accretion disk. Rays corresponding to the former scenario contribute to the black hole shadow, and their respective pixels are colored black. The latter scenario implies that the ray carries a non-zero specific intensity, and the corresponding pixels are assigned colors based on their specific intersection points with the accretion disk. Furthermore, owing to the geometric discontinuity of the torn disk, a fraction of the rays may neither plunge into the black hole nor intersect the disk. Instead, these rays escape to spatial infinity through the gap between the sub-disks; the pixels corresponding to these escaping rays are left white.

Next, to determine the ultimate fate of the light ray corresponding to each pixel $(x,y)$ on the observational screen, it is essential to establish a rigorous mathematical mapping between the screen coordinates and the initial conditions of the photon $(t,r,\theta,\varphi,p_{t},p_{r},p_{\theta},p_{\varphi})$ defined in the local frame of the black hole. Following the methodology outlined in \cite{2016ApJ...820..105P,2016PhRvD..94h4025Y}, the observer's local coordinate frame can be aligned with that of the black hole through a sequence of spatial translations and rotations. These transformations yield the following relations: 
\begin{eqnarray}
x^{\prime} &=& \mathscr{T}\cos\Phi - x\sin\Phi, \label{18} \\
y^{\prime} &=& \mathscr{T}\sin\Phi + x\cos\Phi, \label{19} \\
z^{\prime} &=& \left(r_{\textrm{obs}} - z\right)\cos\Theta + y\sin\Theta, \label{20}
\end{eqnarray}
where the function $\mathscr{T}$ takes the form
\begin{equation}\label{21}
\mathscr{T} = \left(\sqrt{r_{\textrm{obs}}^{2}+a^{2}}-z\right)\sin\Theta-y\cos\Theta.
\end{equation}
Subsequently, the initial spatial coordinates of the photon parameterized in the Boyer-Lindquist system are derived as
\begin{eqnarray}
t &=& 0, \label{22} \\
r &=& \sqrt{\frac{\mathscr{D}+\sqrt{\mathscr{D}^{2}+4a^{2}z^{\prime2}}}{2}}, \label{23} \\
\theta &=& \arccos\left(\frac{z^{\prime}}{r}\right), \label{24} \\
\varphi &=& \textrm{atan}2\left(y^{\prime},x^{\prime}\right), \label{25}
\end{eqnarray}
where $\mathscr{D}$ is expressed as
\begin{equation}\label{26}
\mathscr{D} = x^{\prime2} + y^{\prime2} + z^{\prime2} - a^{2}.
\end{equation}

Evolving forward in time, the light rays strike the observational screen perpendicularly, possessing a three-velocity of $(v_{x},v_{y},v_{z})=(0,0,-1)$ in the local coordinate frame of the observer. By taking the derivative of both sides of equations \eqref{18}--\eqref{20} with respect to the affine parameter $\lambda$ and substituting this photon three-velocity condition, we obtain the initial velocity components of the light ray evaluated in the local frame of the black hole $(x^{\prime},y^{\prime},z^{\prime})$:
\begin{eqnarray}
\dot{x}^{\prime} &=& \sin\Theta\cos\Phi, \label{27} \\
\dot{y}^{\prime} &=& \sin\Theta\sin\Phi, \label{28} \\
\dot{z}^{\prime} &=& \cos\Theta. \label{29}
\end{eqnarray}
Subsequently, by differentiating equations \eqref{23}--\eqref{25} with respect to the affine parameter and substituting the expressions for $(\dot{x}^{\prime},\dot{y}^{\prime},\dot{z}^{\prime})$, we derive the corresponding three-velocity components parameterized in the Boyer-Lindquist coordinates:
\begin{eqnarray}
\dot{r} &=& \frac{r\mathscr{R}\sin\theta\sin\Theta\cos\Psi+\mathscr{R}^{2}\cos\theta\cos\Theta}{\Sigma}, \label{30} \\
\dot{\theta} &=& \frac{\mathscr{R}\cos\theta\sin\Theta\cos\Psi-r\sin\theta\cos\Theta}{\Sigma}, \label{31} \\
\dot{\varphi} &=& -\frac{\sin\Theta\sin\Psi}{\mathscr{R}\sin\theta}, \label{32}
\end{eqnarray}
where $\mathscr{R}$ and $\Psi$ are defined as $\sqrt{r^{2}+a^{2}}$ and $(\varphi-\Phi)$, respectively.

Based on the relationship between the conjugate momenta and velocities given in equations \eqref{7} and \eqref{8}, the momentum components $p_{r}$ and $p_{\theta}$ can be evaluated directly. Subsequently, by rearranging equation \eqref{6}, we obtain the explicit expression for $\dot{t}$:
\begin{equation}\label{33}
\dot{t} =\left(\frac{\Sigma}{\Sigma-2r}\right)\left(-p_{t} - \frac{2ar\sin^{2}\theta}{\Sigma}\dot{\varphi}\right).
\end{equation}
Substituting this expression into the Lagrangian in equation \eqref{5} and imposing the null geodesic constraint $\mathscr{L}=0$, we derive the relationship between $p_{t}$ and the spatial velocity components $(\dot{r},\dot{\theta},\dot{\varphi})$:
\begin{equation}\label{34}
(-p_{t})^{2} = \left(\frac{\Sigma-2r}{\Sigma\Delta}\right)\left(\Sigma\dot{r}^{2}+\Sigma\Delta\dot{\theta}^{2}\right)+\Delta\dot{\varphi}^{2}\sin^{2}\theta.
\end{equation}
By inserting the kinematic results from equations \eqref{30}--\eqref{32} into the equation above, the conserved quantity corresponding to the photon's specific energy, $p_{t}=-E$, is uniquely determined. Furthermore, the other fundamental constant of motion, namely the specific angular momentum $L=p_{\varphi}$, can also be calculated as follows:
\begin{equation}\label{35}
p_{\varphi} = \frac{\left(\Sigma\Delta\dot{\varphi}+2arp_{t}\right)\sin^{2}\theta}{\Sigma-2r}.
\end{equation}

Since the photon frequency does not alter the geometric trajectory of the null geodesic, the derived initial conjugate momenta can be safely renormalized. Specifically, $p_{r}$, $p_{\theta}$, and $p_{\varphi}$ are divided by the specific energy determined from equation \eqref{34}, and $p_{t}$ is subsequently fixed at $-1$ \cite{Cunha:2016bpi}. Consequently, the complete set of initial conditions for the photon, parameterized in the local frame of the black hole, is explicitly specified as $(t,r,\theta,\varphi,-1,p_{r},p_{\theta},p_{\varphi})$. By employing a fifth- and sixth-order Runge-Kutta-Fehlberg (RKF56) integrator with an adaptive step size to numerically integrate the canonical equations \eqref{17}, we can effectively reconstruct the complete trajectory of the light ray corresponding to each pixel $(x,y)$. Notably, because the photon trajectories are traced backward in time from the observational screen, a negative integration step size must be strictly adopted during the numerical integration process.
\section{Results}
The observational field of view is set to $x \in [-15,15]M$ and $y \in [-15,15]M$, with a high resolution of $1500 \times 1500$ pixels. Figure 2 illustrates the inner shadows of a Kerr black hole illuminated by a torn accretion disk under various $r_{\textrm{cut}}$ configurations. When the inner and outer sub-disks connect at $r_{\textrm{cut}}=r_{\textrm{isco}}$\footnote{Here, $r_{\text{isco}}$ denotes the radius of the innermost stable circular orbit (ISCO) for a timelike particle. Given our fixed dimensionless spin parameter of $a = 0.94$, this critical radius is determined to be $r_{\text{isco}} \approx 2.024M$.}, the center of the image prominently displays the inner shadow contributed by the event horizon. As the inclination angle $\sigma$ of the outer sub-disk increases, this inner shadow undergoes progressive erosion. This erosion occurs because the gradually elevated outer disk intercepts photon trajectories that would have otherwise plunged into the black hole. Surrounding the inner shadow, a bright ring primarily contributed by the inner sub-disk is discernible. The thickness and morphology of this bright ring are highly dependent upon the tearing radius and the tilt angle of the outer disk, respectively. Interestingly, beyond the bright ring, we identify crescent-shaped shadows that expand in correlation with an increasing outer disk tilt angle. This phenomenon is attributed to the spatial discontinuity between the inner and outer sub-disks, which creates a geometric window for photons to plunge into the black hole. As the tearing radius increases outward (from left to right), these crescent-shaped shadows are progressively disrupted and ultimately vanish when $r_{\textrm{cut}}$ becomes sufficiently large. Concurrently, this morphological transition is accompanied by the radiation flux from the inner sub-disk gradually dominating the overall emission. Furthermore, within specific parameter spaces, we detect escaping rays, denoted by white pixels. This feature is similarly facilitated by the geometric gap within the torn disk, which provides a collision-free pathway for photons to propagate from the observational screen to spatial infinity without intersecting either the accretion disk or the black hole.
\begin{figure*}
\center{
\includegraphics[width=15cm]{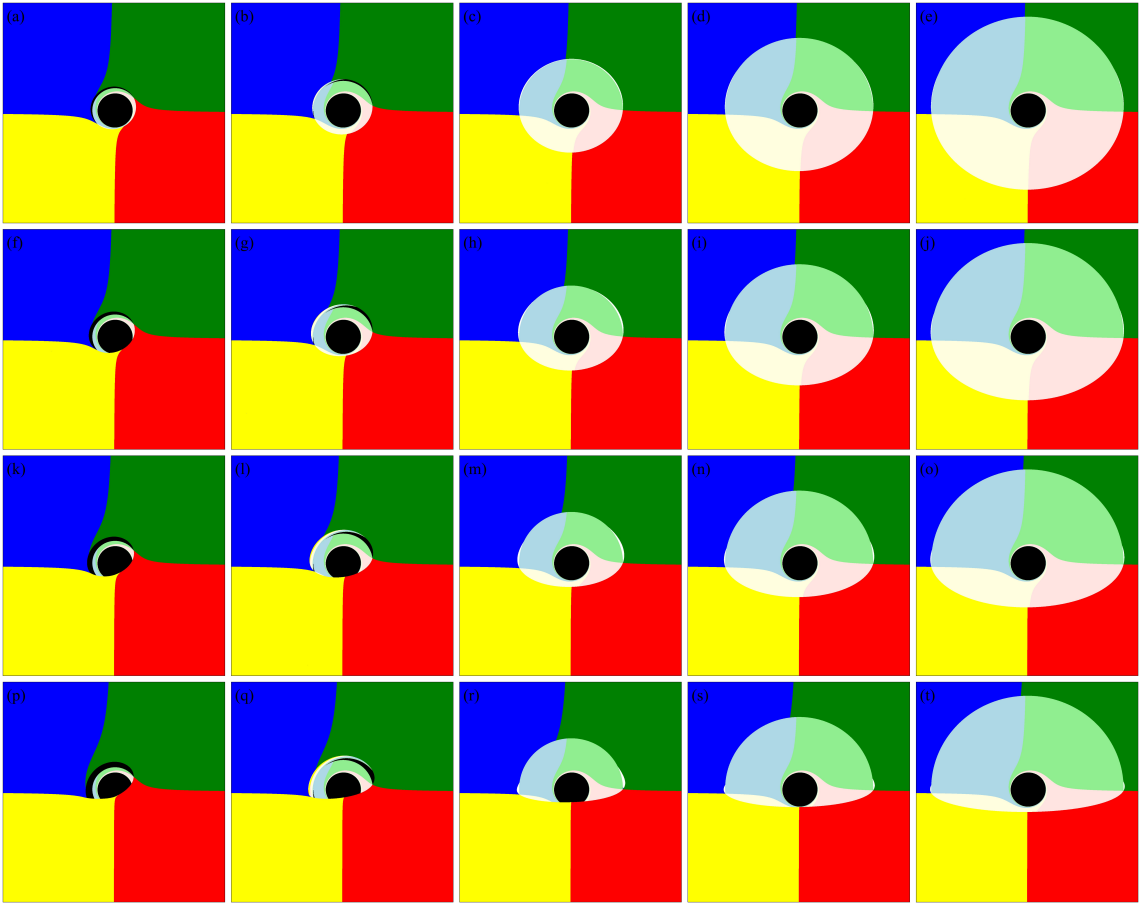}
\caption{Inner shadows of a Kerr black hole illuminated by a torn accretion disk across different parameter spaces. From top to bottom, the inclination angle of the outer tilted sub-disk is set to $15^{\circ}$, $30^{\circ}$, $45^{\circ}$, and $60^{\circ}$, respectively. From left to right, the inner and outer sub-disks connect at $r_{\textrm{cut}}=r_{\textrm{isco}}$, $3$, $6$, $9$, and $12$. The viewing inclination and observer azimuthal angle are fixed at $\Theta=17^{\circ}$ and $\Phi=0^{\circ}$, respectively. Notably, eyebrow-like shadows emerge, and the erosion of the inner shadow becomes evident within specific parameter spaces.}} \label{fig2}
\end{figure*}

By increasing the viewing inclination to $\Theta=50^{\circ}$, we repeat the simulations analogous to those in figure 2, with the corresponding results presented in figure 3. Initially, in the top two rows, we identify phenomena consistent with those observed in figure 2. Furthermore, when the line of sight lies below the outer accretion disk (the bottom two rows), the inner shadow exhibits an inverted morphology. Concurrently, the inner shadow can be bifurcated by a bright emission band originating from the inner sub-disk, as explicitly shown in panels (k), (l), (p), and (q). As the disk tearing radius shifts outward, the black hole shadow transitions into a composite structure comprising two parts with drastically disparate sizes: the larger component resembles an arch, while the smaller one is as thin as an eyebrow. Ultimately, when the disk tearing occurs at a sufficiently large distance, the black hole shadow coalesces into a singular image.
\begin{figure*}
\center{
\includegraphics[width=15cm]{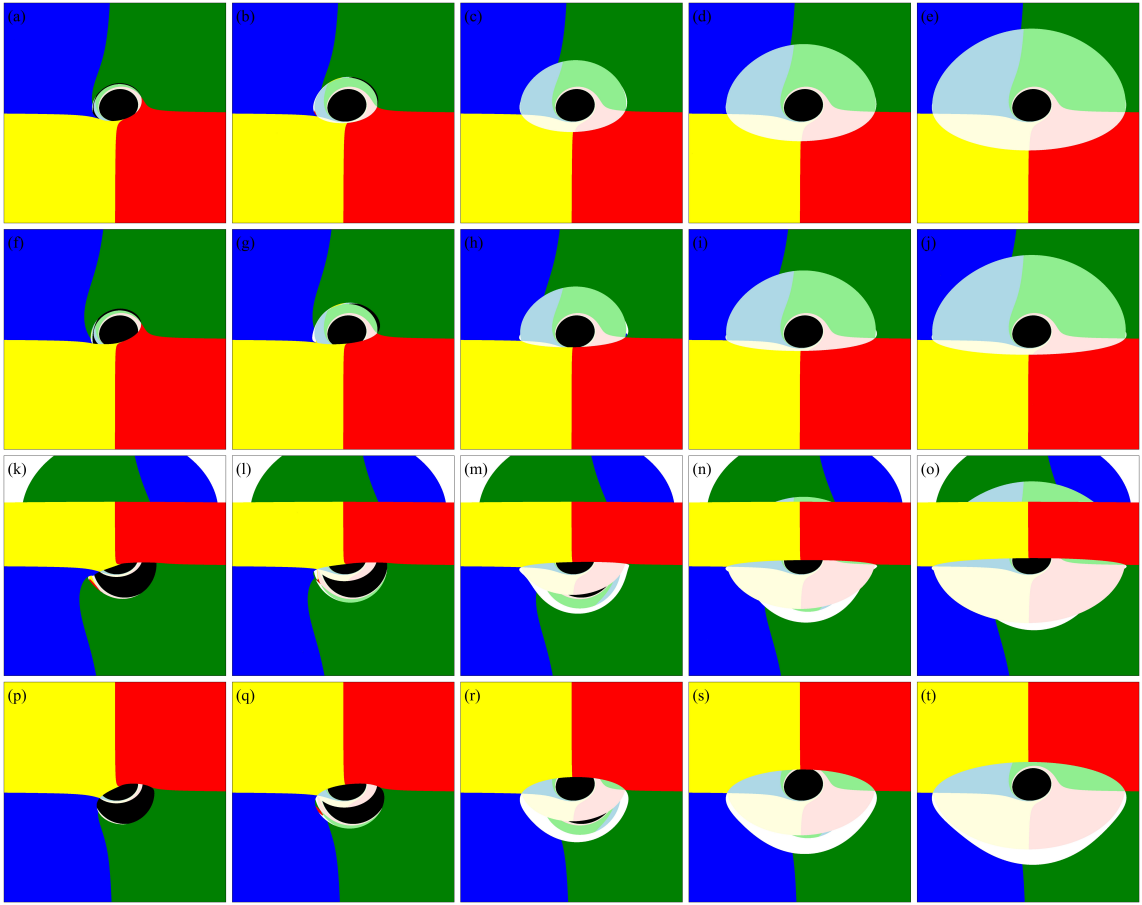}
\caption{Similar to figure 2, but for a viewing inclination of $\Theta=50^{\circ}$.}} \label{fig3}
\end{figure*}

When the viewing inclination is further elevated to $\Theta=80^{\circ}$, our simulation results are displayed in figure 4. It is evident that when the disk tearing radius is proximal to the $r_{\textrm{isco}}$ (the left two columns), the bright emission band contributed by the inner sub-disk partitions the black hole shadow into two distinct regions: a leaf-shaped upper segment and a nearly semicircular lower segment. Meanwhile, as the tilt angle of the outer sub-disk increases, the upper shadow progressively expands in size, whereas the morphology of the lower shadow exhibits a tendency to converge toward the $y$-axis. As the tearing radius is enlarged, these two shadow components gradually separate from each other, accompanied by a noticeable shrinkage of the lower shadow.
\begin{figure*}
\center{
\includegraphics[width=15cm]{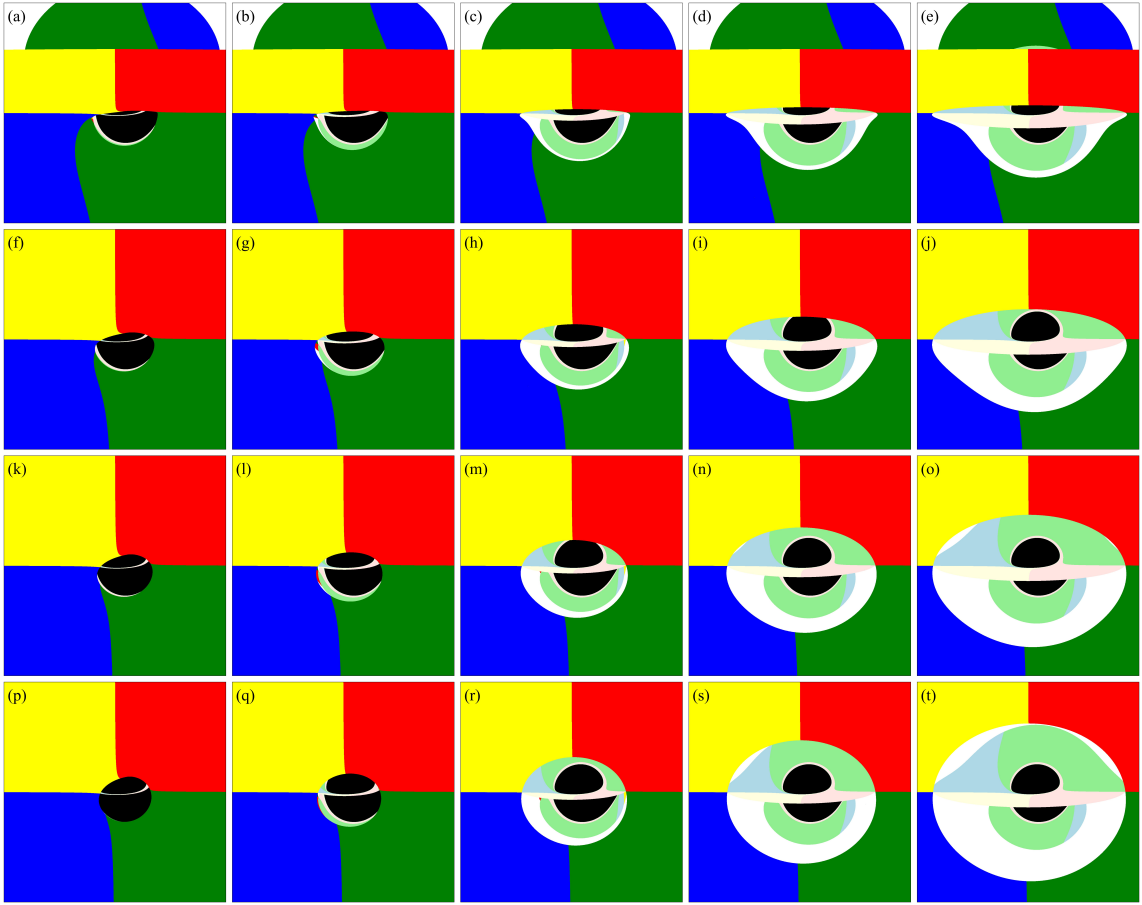}
\caption{Similar to figure 2, but for a viewing inclination of $\Theta=80^{\circ}$.}} \label{fig4}
\end{figure*}

Synthesizing the results from figures 2 through 4, we assert that the torn accretion disk possesses the profound capability to modify the intrinsic black hole shadow. Across the majority of the parameter space, the black hole shadow manifests as two non-negligible components, each exhibiting a rich variety of morphologies, such as crescent, arched, and eyebrow-like shapes. It is worth emphasizing that in purely equatorial or simply tilted accretion environments, it is exceedingly difficult for a Kerr black hole to generate such a double-shadow structure. On the other hand, despite the potential for scalar hairy black holes to produce multi-shadow structures \cite{2015PhRvL.115u1102C,Cunha:2016bpi,2016PhRvD..94j4023C,2021PhRvD.104h4021W}, there remains a stark morphological contrast between those theoretical predictions and the images found in our present study. Therefore, the novel shadow structures identified in this work have significant potential to serve as robust probes for torn accretion environments.
\begin{figure*}
\center{
\includegraphics[width=15cm]{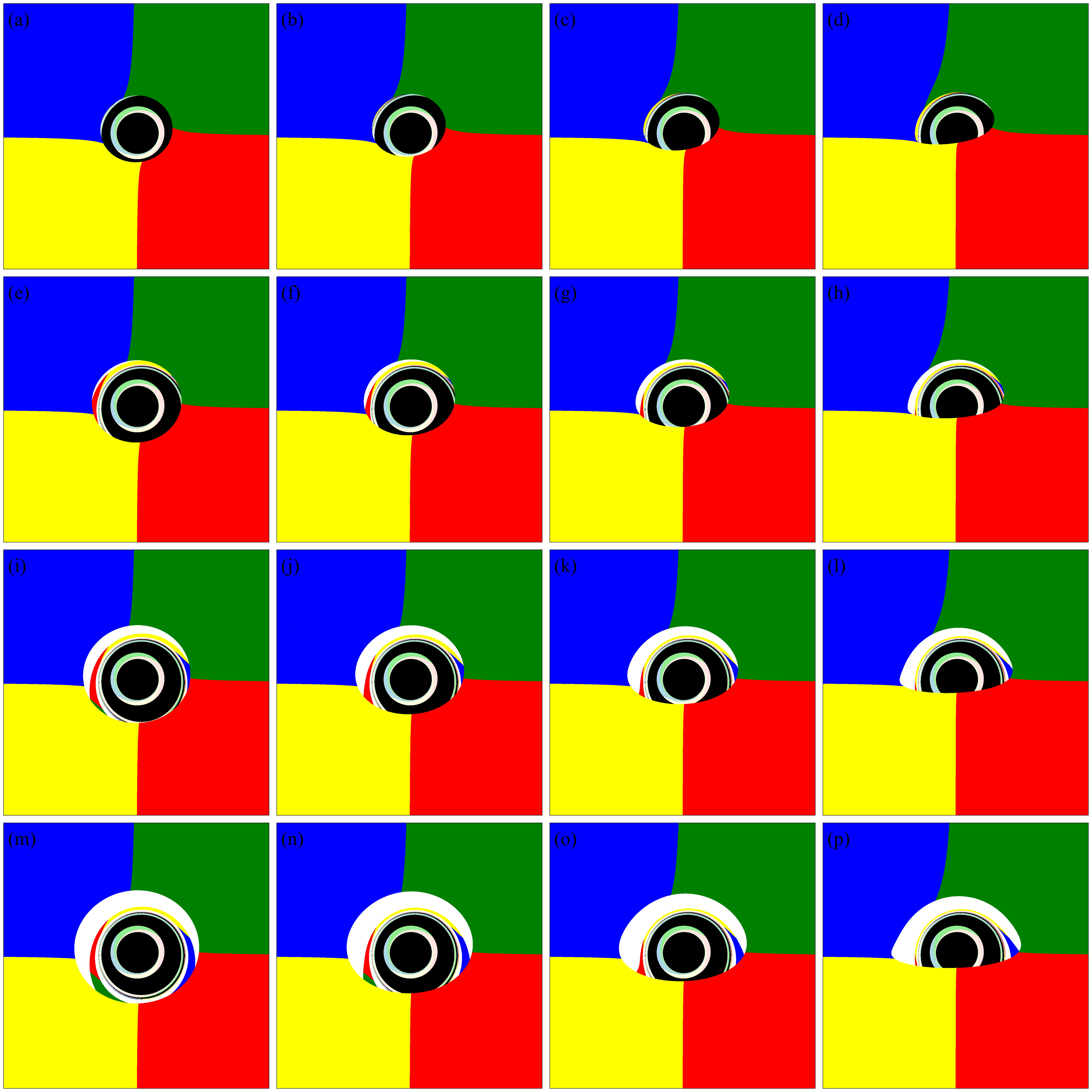}
\caption{Inner shadows of a Kerr black hole illuminated by a torn accretion disk across different parameter spaces. Here, the inner and outer accretion disks are radially discontinuous. The inner accretion sub-disk is strictly confined to the equatorial plane, with its outer boundary fixed at the ISCO ($r^{\textrm{inner}}_{\textrm{out}}=r_{\textrm{isco}}$). From top to bottom, the inner boundary of the outer accretion sub-disk is located at $r^{\textrm{outer}}_{\textrm{in}}=3$, $4$, $5$, and $6$, respectively. From left to right, the inclination angle of the outer accretion sub-disk is set to $15^{\circ}$, $30^{\circ}$, $45^{\circ}$, and $60^{\circ}$. The viewing inclination and observer azimuthal angle are fixed at $\Theta=17^{\circ}$ and $\Phi=0^{\circ}$, respectively.}} \label{fig5}
\end{figure*}

Next, we fix $r^{\textrm{inner}}_{\textrm{out}}=r_{\textrm{isco}}$ to investigate the influence of $r^{\textrm{outer}}_{\textrm{in}}$ on the inner shadow of the Kerr black hole across different parameter spaces. Figure 5 illustrates the combined effects of the outer disk's tilt angle and its inner boundary location on the black hole shadow at a viewing inclination of $\Theta=17^{\circ}$. We observe that the black hole shadow is composed of multiple distinct components: a nearly circular inner shadow located at the center of the image, a prominent primary shadow ring, and an extremely thin secondary shadow ring attached to its periphery. These individual components are spatially isolated from one another by bright rings contributed by the inner accretion sub-disk. Furthermore, apart from being obscured by the tilted outer disk within specific parameter spaces, the fundamental size of the central inner shadow remains virtually unaffected. Concurrently, the shadow rings are highly susceptible to erosion by the surrounding accretion environment. With the exception of panels (a), (e), (i), (j), (m), and (n) where complete ring structures are identifiable, the shadow rings predominantly manifest as semicircular or three-quarter circular arcs. The underlying formation mechanism of these shadow rings dictates that the spatial gap between the inner and outer sub-disks permits a fraction of photons to bypass collisions with the accretion flow and plunge directly into the event horizon. Meanwhile, the secondary shadow ring is generated by photons that execute an additional orbit around the black hole prior to plunging. It is anticipated that higher-order shadow rings theoretically exist, and as the order increases, their contours asymptotically approach the critical curve.

Alternatively, the formation of these shadow rings can be comprehended from another perspective. If the equatorial inner accretion disk were entirely absent (i.e., no accreting matter in the region $r < r^{\textrm{outer}}_{\textrm{in}}$), a complete, quasi-circular shadow would be expected. The introduction of the inner accretion disk superimposes bright emission rings onto this otherwise intact shadow, thereby inducing the emergence of the shadow rings. Evidently, these bright rings can manifest in multiple higher-order iterations, with each successive order systematically dissecting the shadow and yielding corresponding shadow rings. In summary, the spatial separation between the inner and outer sub-disks resulting from the tearing event establishes the essential prerequisite for the formation of shadow rings, while the inclination of the outer disk facilitates the erosion of both the shadow rings and the central inner shadow. Consequently, the exotic shadow features unveiled in figure 5 can serve as a robust diagnostic tool for probing the tearing behavior of accretion disks surrounding black holes. 
\begin{figure*}
\center{
\includegraphics[width=15cm]{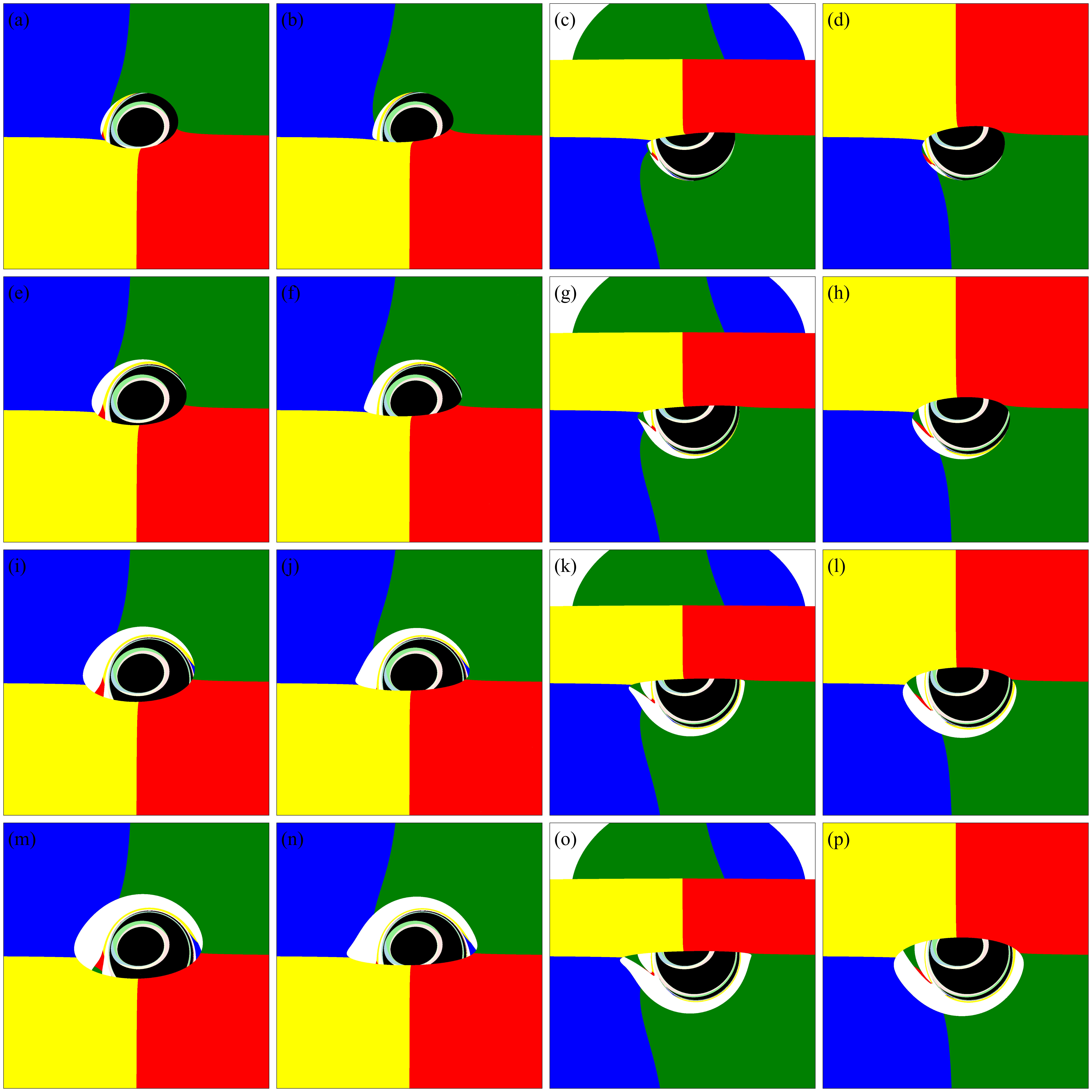}
\caption{Similar to figure 5, but for a viewing inclination of $\Theta=50^{\circ}$.}} \label{fig6}
\end{figure*}

Modifying the viewing inclination to $\Theta=50^{\circ}$, we repeat the simulations from figure 5, with the corresponding results presented in figure 6. First, we observe that the rightmost two columns exhibit an inverted morphology, which originates from the line of sight being situated below the outer tilted accretion disk. Second, compared to figure 5, the central inner shadow undergoes noticeable deformation. This is attributed to the increased viewing inclination, which stretches the inner shadow horizontally and compresses it vertically. Furthermore, instances occur where the inner shadow is partially obscured by the surrounding accretion environment. Subsequently, it is evident that the bright ring enveloping the inner shadow is similarly contributed by the inner accretion sub-disk. However, due to obscuration by the outer tilted disk, this bright ring appears intact only within a minority of the parameter space, such as in the left columns; in other cases, it manifests as banded structures. Beyond this bright ring (or band) lies an irregular shadow, which morphs from the primary shadow ring identified in figure 5. Moreover, driven by the frame-dragging effect, this shadow structure is distinctly shifted to the right. Finally, we find that the secondary shadow ring from figure 5 transitions into an elongated, non-negligible shadow band here, predominantly featured in the right two columns.
\begin{figure*}
\center{
\includegraphics[width=15cm]{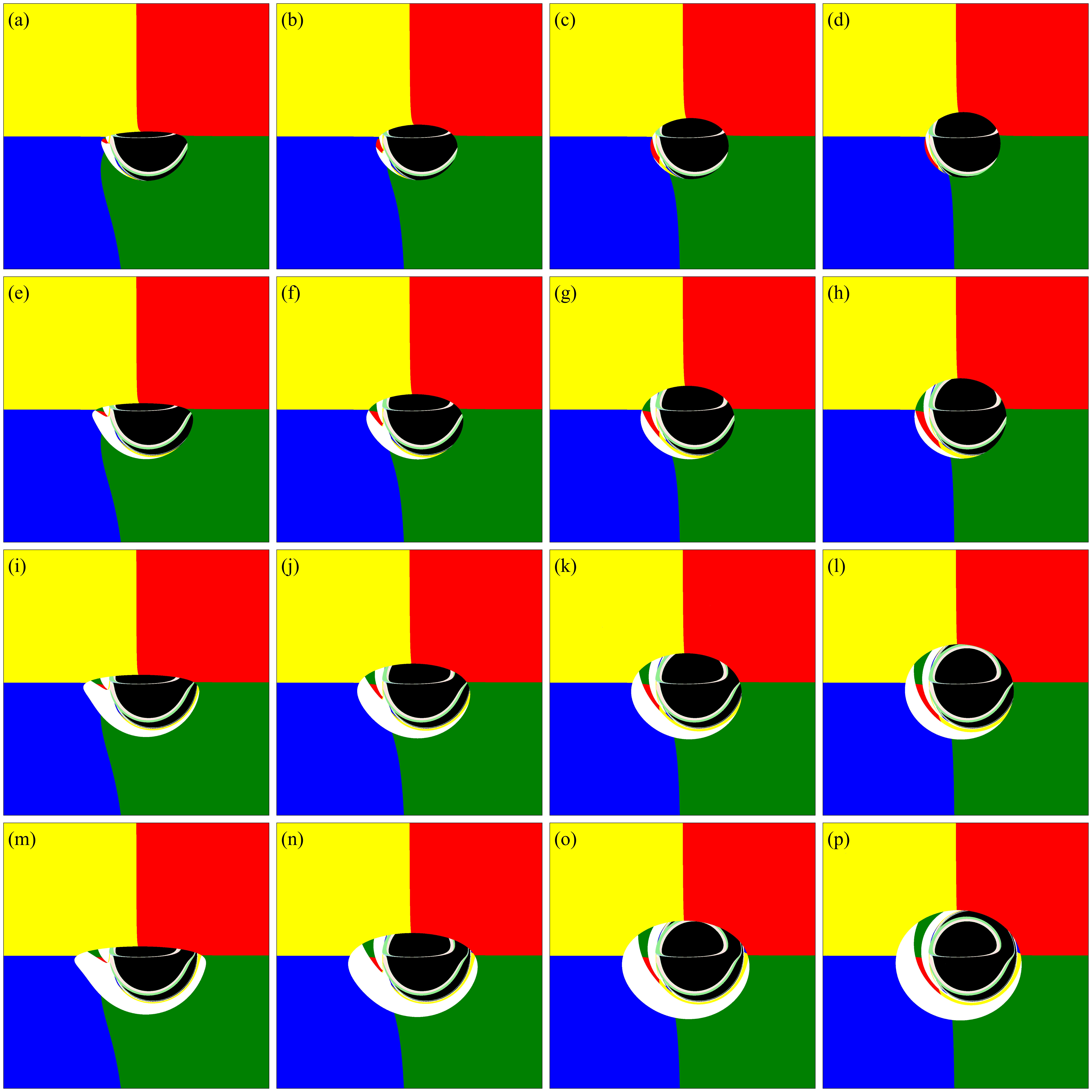}
\caption{Similar to figure 5, but for a viewing inclination of $\Theta=80^{\circ}$.}} \label{fig7}
\end{figure*}

Further elevating the viewing inclination to $\Theta=80^{\circ}$ yields the results shown in figure 7. We find that the structure of the black hole shadow becomes exceedingly complex. Both increasing the tilt angle of the outer accretion disk and enlarging $r^{\textrm{outer}}_{\textrm{in}}$ lead to a pronounced expansion of the overall shadow region. This expansion occurs because both geometric adjustments provide a spatial window for photons to evade collisions with the accretion disk, enabling them to plunge smoothly into the black hole and thereby contribute to the inner shadow. By examining panel (p), we can comprehend the formation mechanism of these exotic shadows. Theoretically, when observing a Kerr black hole with a spin parameter of $a=0.94$ from a near edge-on perspective (without accretion flow), the standard shadow exhibits a ``D-like'' morphology. However, the equatorial sub-disk, clinging to the event horizon, contributes multiple orders of bright rings within the critical curve. Concurrently, the tilted accretion disk intercepts a portion of the photons that would have otherwise plunged into the black hole, effectively eroding the shadow. The synergistic interplay of these two effects produces the singular structure observed in panel (p). If the bright rings (or bands) were artificially disregarded, the underlying D-like shadow could still be conceptually visualized.

\begin{figure*}
\center{
\includegraphics[width=15cm]{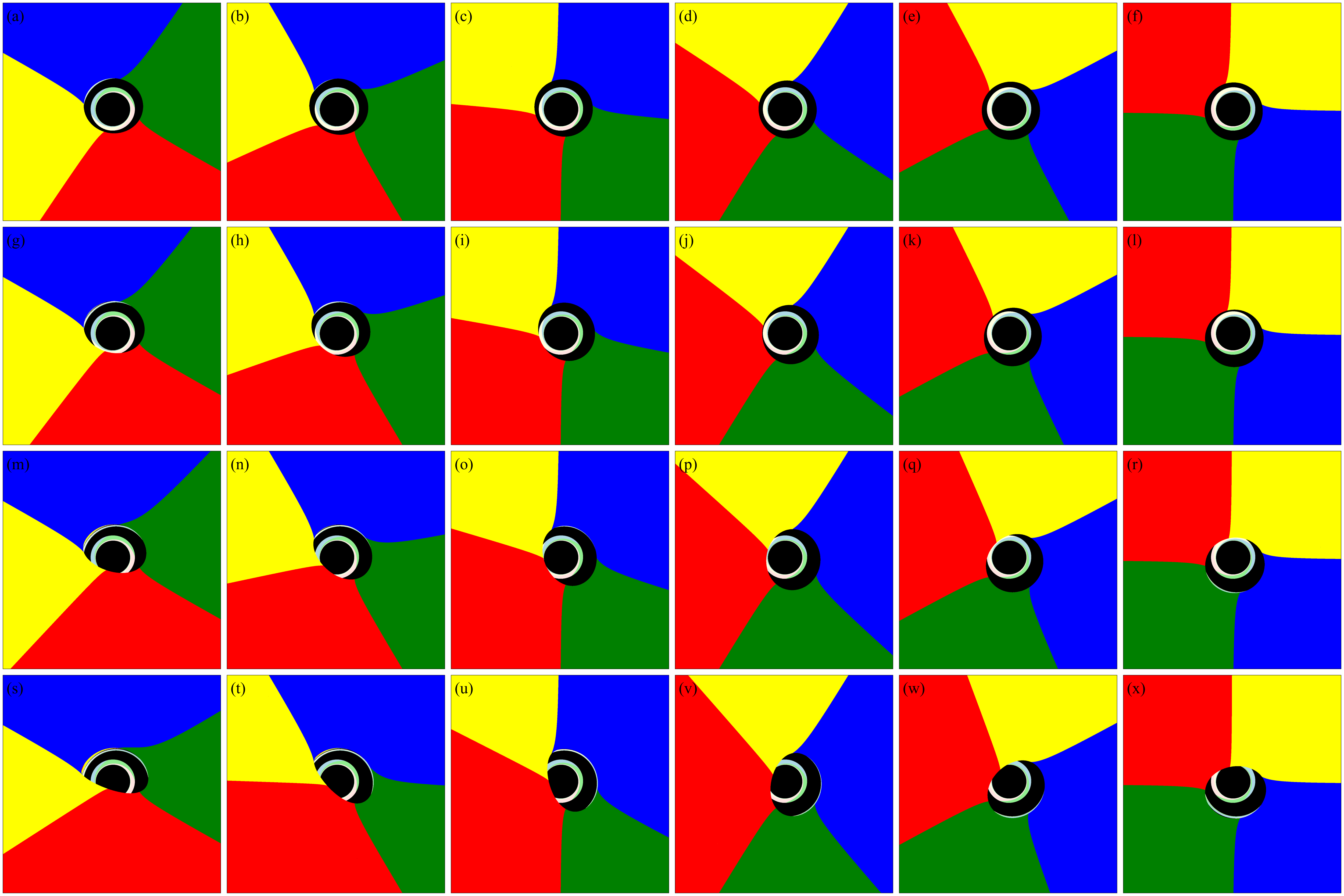}
\caption{Inner shadows of a Kerr black hole illuminated by a torn accretion disk across different parameter spaces. From top to bottom, the inclination angle of the outer tilted sub-disk is set to $15^{\circ}$, $30^{\circ}$, $45^{\circ}$, and $60^{\circ}$, respectively. From left to right, the observer's azimuthal angle $\Phi$ increases from $30^{\circ}$ to $180^{\circ}$ in increments of $30^{\circ}$. Here, the inner accretion sub-disk spans from the event horizon to the ISCO. The inner boundary of the outer sub-disk is fixed at $r^{\textrm{outer}}_{\textrm{in}} =3$, and the viewing inclination is set to $\Theta=17^{\circ}$.}} \label{fig8}
\end{figure*}
Owing to the tilt of the outer accretion sub-disk, it is anticipated that the present model no longer preserves the axisymmetry inherent in purely equatorial accretion disk configurations. Consequently, it is imperative to investigate the variations in the resulting images induced by altering the observer's azimuthal angle, $\Phi$. Figure 8 illustrates the black hole shadows perceived by an observer at a fixed viewing inclination of $\Theta=17^{\circ}$ across various azimuthal angles. We observe that when the inclination angle of the outer tilted sub-disk is $15^{\circ}$, shifting the observer counterclockwise (i.e., increasing the azimuthal angle $\Phi$) induces a global rotation of the entire image. In this low-tilt regime, the black hole shadow is predominantly composed of an intact inner shadow and complete shadow rings. 

However, upon significantly increasing the tilt angle of the outer sub-disk, we identify that the structural integrity of both the inner shadow and the shadow rings is conspicuously broken. This disruption occurs because the elevated outer sub-disk visually obscures the photon trajectories. It is readily apparent that as the disk inclination angle increases (from top to bottom), the erosion experienced by both the shadow rings and the inner shadow becomes progressively more severe. In these highly tilted cases, varying the azimuthal angle continues to dictate the rotation of the global shadow structure while simultaneously modulating the specific obscuration and erosion effects. 

\begin{figure*}
\center{
\includegraphics[width=15cm]{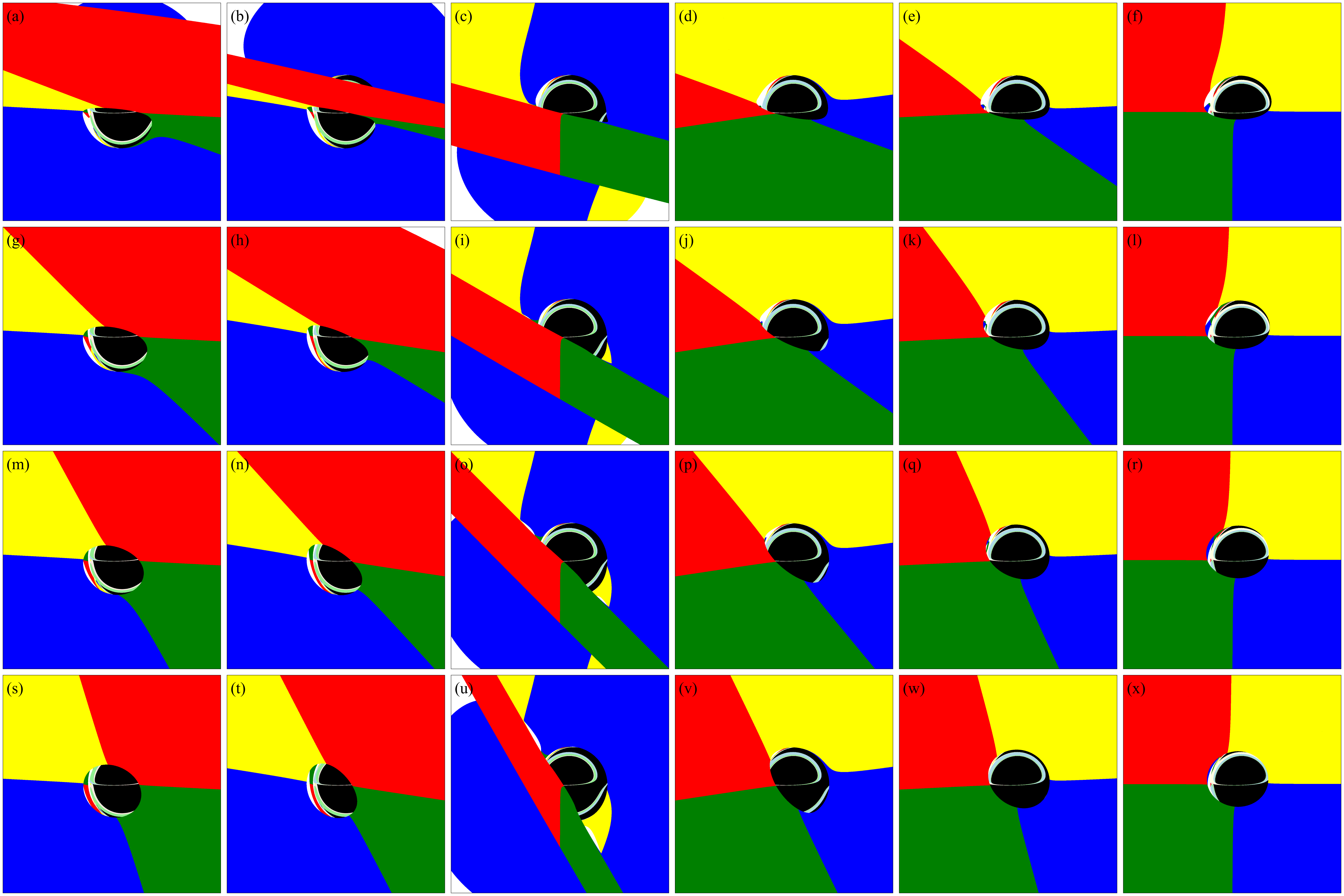}
\caption{Similar to figure 8, but for a viewing inclination of $\Theta=85^{\circ}$.}} \label{fig9}
\end{figure*}
Elevating the viewing inclination to an extreme $\Theta=85^{\circ}$, we further investigate the influence of the observer's azimuthal angle on the black hole shadow. As depicted in figure 9, scanning from left to right across each row, we consistently observe a shadow inversion process. This phenomenon occurs because, as the observer rotates, the line of sight progressively transitions from beneath the outer tilted sub-disk to above it. Concurrently, we discover that the black hole shadow attains its maximum apparent size at an azimuthal angle of $\Phi=90^{\circ}$ (the third column from the left). This maximization arises because, under this specific configuration, the tilted outer sub-disk appears almost completely edge-on to the observer. Consequently, its direct geometric capacity to intercept photon trajectories is severely minimized, leaving it to obscure rays solely through extreme gravitational lensing and the frame-dragging effect. This minimal geometric cross-section effectively opens a substantially larger spatial window, permitting an increased number of photons to plunge into the black hole and thereby enlarging the shadow. It is worth noting that adjusting the observer's azimuthal angle can be kinematically conceptualized as a physical rotation of the accretion disk itself. Therefore, figures 8 and 9 can be effectively regarded as a phenomenological representation of the dynamical evolution of the black hole shadow.
\section{Conclusions and Discussions}
In this paper, adopting a purely geometric perspective, we constructed a torn accretion disk model composed of an aligned inner equatorial sub-disk and a misaligned outer tilted sub-disk to systematically investigate its imprint on the morphological evolution of the inner shadow of a Kerr black hole ($a=0.94$). By extensively exploring the parameter space, we demonstrated that when the inner and outer sub-disks are radially continuous, the elevation of the tilted outer disk directly intercepts photon trajectories that would have otherwise plunged into the black hole. This geometric effect leads to severe erosion of the central inner shadow. Concurrently, the tearing radius creates a geometric window for photons, inducing the emergence of crescent-shaped shadows that expand with increasing tilt angles. Notably, at higher viewing inclinations, the inner shadow can exhibit an inverted morphology and become bifurcated by the bright emission band from the inner disk, resulting in a double-shadow structure with drastically disparate component sizes.

Furthermore, upon introducing a radial discontinuity $(r^{\textrm{outer}}_{\textrm{in}} > r^{\textrm{inner}}_{\textrm{out}})$, the black hole shadow manifests a highly characteristic multi-ring architecture, explicitly featuring primary and secondary shadow rings. The essence of these shadow rings lies in the introduction of the inner accretion sub-disk, which effectively dissects the otherwise intact, quasi-circular shadow through multiple orders of bright emission rings. 

More fundamentally, we emphasize that within the framework of a torn accretion disk, the observed black hole silhouette does not manifest exclusively as a traditional shadow (contributed by the critical curve) or an inner shadow, but rather emerges as a hybrid coexistence of both. In a scenario featuring a purely equatorial accretion disk that extends seamlessly down to the event horizon, the black hole image is strictly dominated by the inner shadow. However, elevating and tilting the outer portion of this disk systematically intercepts a fraction of the photon trajectories that would otherwise construct the inner shadow, thereby causing its prominent erosion. Concurrently, this geometric reconfiguration opens spatial windows that permit a different set of photons to plunge directly into the black hole. These newly plunging photons effectively delineate specific segments of the traditional black hole shadow. It is precisely this simultaneous generation and spatial superposition of both shadow types that gives rise to the diverse and exotic shadow morphologies discovered in this work.

In conclusion, the torn accretion environment possesses a profound capability to reshape the intrinsic geometry of black hole shadows. The novel structures unveiled in this work, such as double shadows and shadow rings, are virtually impossible to generate within traditional single-equatorial or purely tilted disk paradigms. Furthermore, although it is known that certain exotic spacetimes---such as black holes endowed with scalar hair \cite{2015PhRvL.115u1102C,Cunha:2016bpi,2016PhRvD..94j4023C,2021PhRvD.104h4021W}, as well as binary and triple black hole systems \cite{2011PhRvD..84f3008N,2012PhRvD..86j3001Y,2015CQGra..32f5002B,2018PhRvD..98d4053C,2025EPJC...85..905L}---can also exhibit eyebrow-like shadows, and the latter two may even generate shadow rings depending on their spatial configurations, their resulting visual signatures remain morphologically distinct from the accretion-driven structures discovered herein. Therefore, these distinctive observational signatures not only enrich the theoretical template library of black hole images but also hold great promise as robust diagnostic tools for probing the dynamical tearing behavior of accretion disks in the strong-field regime with the next-generation Event Horizon Telescope (ngEHT).

It is necessary to point out that the current work does not treat the accreting matter as a physical plasma with associated radiative and kinematic properties; rather, we phenomenologically treat it as a static geometric emission proxy. During the ray-tracing process, if a photon trajectory intersects the accretion matter, it is deemed emissive and mapped to a corresponding color on the observer's screen. Such a simplified binary procedure might inevitably filter out certain micro-structural features. For instance, authentic accreting matter possesses a distinct four-velocity profile, which subjects the local specific intensity to significant Doppler effects. Consequently, in the plunging region near the black hole, although the matter is intrinsically emissive, extreme gravitational and kinematic redshifts could render the corresponding flux on the observer's screen exceedingly dim. In other words, if the fully relativistic kinematics of the accretion flow were incorporated, the bright rings inside the shadow or other fine emission features obtained herein might become less discernible. On the other hand, the inner and outer sub-disks resulting from the tearing process would likely exhibit distinct thermodynamic properties (e.g., varying temperature and pressure gradients). Thus, the realistic emission would not be governed merely by the simplistic ``0--1'' step-function employed here. It is reasonable to anticipate that the incorporation of a comprehensive radiative model would yield much richer substructures within the four-color regions of our present images. Crucially, however, the photon trajectories that ultimately define the black hole shadow are precisely those that never intersect the accreting matter. Therefore, even when subjected to a more realistic astrophysical accretion environment, the core phenomena discovered in this work---such as the shadow erosion, tearing, rotation, and deformation---will robustly survive. In the near future, we plan to construct a physically motivated torn accretion disk system within an astrophysical context, coupling it with general relativistic radiative transfer to further investigate the observational signatures of black holes embedded in such environments.

\acknowledgments
The authors are very grateful to the referee for insightful comments and valuable suggestions. This research has been supported by the National Natural Science Foundation of China [Grant No. 12403081].

\bibliographystyle{JHEP}
\bibliography{references}

\end{document}